\documentclass[namedreferences]{SolarPhysics}
\usepackage[optionalrh,natbib]{spr-sola-addons} 
\usepackage{graphicx}                    
\usepackage{amssymb}                     
\usepackage{color}                       
\usepackage{url}                         
\usepackage{enumitem} 
\usepackage{multirow}
\usepackage{booktabs}

\newcommand{\adv}{    {\it Adv. Space Res.}}

\newcommand{\apj}{    {\it Astrophys. J.}}
\newcommand{\apjl}{   {\it Astrophys. J. Lett.}}

\newcommand{\grl}{    {\it Geophys. Res. Lett.}}

\newcommand{\jgr}{    {\it J. Geophys. Res.}}

\newcommand{\solphys}{{\it Solar Phys.}}

\newcommand{\ssr}{    {\it Space Sci. Rev.}}

\begin{document}

\begin{article}

\begin{opening}

\title{Assessing the Constrained Harmonic Mean Method for Deriving the Kinematics of ICMEs with a Numerical Simulation}

%
\author{T.~\surname{Rollett}$^{1}$\sep
        M.~\surname{Temmer}$^{1}$\sep  
        C.~\surname{M\"ostl}$^{1,2}$\sep      
        N.~\surname{Lugaz}$^{3}$\sep  
        A.M.~\surname{Veronig}$^{1}$\sep              
        U.V.~\surname{M\"ostl}$^{1}$}

%
\runningauthor{T.~Rollett \textit{et al.}}
\runningtitle{Assessing the CHM Method with a Numerical Simulation}

%
\institute{$^{1}$ IGAM-Kanzelh\"ohe Observatory, Institute of Physics, University of Graz, Graz, A-8010, Austria
                  email: \url{tanja.rollett@uni-graz.at}\\ 
           $^{2}$ Space Sciences Laboratory, University of California, Berkeley, CA 94720-7450, USA\\
           $^{3}$ Space Science Center, University of New Hampshire, Durham, New Hampshire, USA\\}

\begin{abstract}
In this study we use a numerical simulation of an artificial coronal mass ejection (CME) to validate a method for calculating propagation directions and kinematical profiles of interplanetary CMEs (ICMEs). In this method observations from heliospheric images are constrained with \textit{in-situ} plasma and field data at 1~AU. These data are used to convert measured ICME elongations into distance by applying the Harmonic Mean approach that assumes a spherical shape of the ICME front. We use synthetic white-light images, similar as observed by STEREO-A/HI, for three different separation angles between remote and \textit{in-situ} spacecraft, of $30^\circ$, $60^\circ$, and $90^\circ$. To validate the results of the method they are compared to the apex speed profile of the modeled ICME, as obtained from a top view. This profile reflects the ``true'' apex kinematics since it is not affected by scattering or projection effects. In this way it is possible to determine the accuracy of the method for revealing ICME propagation directions and kinematics. We found that the direction obtained by the constrained Harmonic Mean method is not very sensitive to the separation angle ($30^\circ$ sep: $\phi=$ W7; $60^\circ$ sep: $\phi=$ W12; $90^\circ$ sep: $\phi=$ W15; true dir.: E0/W0). For all three cases the derived kinematics are in a relatively good agreement with the real kinematics. The best consistency is obtained for the $30^\circ$ case, while with growing separation angle the ICME speed at 1~AU is increasingly overestimated ($30^\circ$ sep: $\Delta V_{\rm arr}\approx-50$ km s$^{-1}$, $60^\circ$ sep: $\Delta V_{\rm arr}\approx+75$ km s$^{-1}$, $90^\circ$ sep: $\Delta V_{\rm arr}\approx+125$ km s$^{-1}$). Especially for future L$_4$/L$_5$ missions the $60^\circ$ separation case is highly interesting in order to improve space weather forecasts.
\end{abstract}

%

\end{opening}

%
\section{Introduction}
\label{s:intro}

Coronal Mass Ejections (CMEs) are the most powerful eruptions on the Sun and are currently observed by a relatively large number of instruments. Space-based coronagraphs are located at three different positions to monitor the solar corona from all sides. \textit{SOlar and Heliospheric Observatory} (SOHO) LASCO coronagraphs \citep[C2 and C3:][]{bru95} at L$_1$ provide white-light images up to a distance of 32 R$_{\odot}$. The two spacecraft of the \textit{Solar TErrestrial RElations Observatory} \citep[STEREO:][]{kai08} move apart by $45^\circ$ per year and are currently at a position of more than $120^\circ$ eastern and western from Earth, respectively. As part of the \textit{Sun-Earth Connection Coronal and Heliospheric Investigation} suite \citep[SECCHI:][]{how08,har09} the COR1 and COR2 coronagraphs aboard the two spacecraft STEREO-A and STEREO-B provide continuous white-light observations up to 15 R$_{\odot}$.
Combining coronagraphs with the Heliospheric Imagers HI1 and HI2 \citep{eyl09}, which are also part of the SECCHI suite aboard STEREO, the entire propagation of an ICME from Sun to 1~AU is covered. Both HIs are off-pointed from the center of the Sun and cover a field of view (FoV) of $4^\circ$ to $24^\circ$ and $18.7^\circ$ to $88.7^\circ$ elongation, respectively.

Such wide FoVs make the interpretation of extended structures in these images rather difficult and Thomson scattering effects \citep{bil66,hun93} as well as effects due to line of sight (LoS) integration have to be taken into account. Methods were developed in order to investigate the propagation of CMEs beyond the typical FoV of coronagraph images, which are up to about $4^\circ$ (COR2) or $9^\circ$ (LASCO C3) elongation. These methods are based on specific geometrical assumptions about the shape of the ICME front. Some assume a point-like structure as the Fixed-$\phi$ method \citep[F$\phi$:][]{she99,rou08} or the triangulation method developed by \citet{liu10a} that uses stereoscopic observations. The Harmonic Mean method \citep{lug10} and the Self Similar Expansion method \citep{dav12} assume a circular front. Fixed-$\phi$, Harmonic Mean \citep{lug10,moe11} and the Self Similar Expansion method \citep{moedav12} are also used as fitting methods by assuming constant ICME speed and direction. In this way, these methods can be used to forecast the ICME arrival at 1~AU. An alternative forecasting tool is the drag based model, developed by \citet{vrs12}. It is based on the idea that the aerodynamic drag in the ambient medium is responsible for the adjustment of the ICME speed to the solar wind speed.

ICMEs are not only remotely observed but also measured \textit{in-situ}, among others, from both STEREO spacecraft \citep[IMPACT:][]{luh08}, from \textit{Wind} \citep{ogi95} and the \textit{Advanced Composition Explorer} \citep[ACE:][]{sto98}. ACE and \textit{Wind} are both located at the L$_1$ point. In order to guard against misunderstandings, we use the term ICME as defined by \citet{rou11}, \textit{i.e.}\ an ICME includes the shock (if present), the dense sheath region and the magnetic flux rope. Furthermore we use this term for white-light observations beyond the coronagraphic FoV as well as for the \textit{in-situ} structure. Using plasma and magnetic field data from \textit{in-situ} observations, in addition to remote sensing of ICMEs, we are able to constrain the results derived from measurements of white-light images \citep{moe09apj,moe10,rol12}. In the following we refer to the method as ``constrained'' Harmonic Mean method (CHM). 
The kinematical profiles resulting from the CHM method may also be compared to the results of the drag based model and serve for studies of the interaction of ICMEs with solar wind strucures \citep{tem11,tem12}.

MHD simulations of ICMEs evolving in interplanetary (IP) space have already been used to validate the Fixed-$\phi$ fitting and Harmonic Mean fitting methods \citep{lug11}. By using synthetically derived ICME density images and time-elongation plots \citep[Jmaps:][]{dav09b}, measured from different viewing angles, we aim to validate the CHM method and present a detailed error analysis in order to improve the forecasting of Earth-directed ICMEs. 
We construct the Jmaps for different separation angles between the remote and the \textit{in-situ} spacecraft observing the ICME, namely $30^\circ$, $60^\circ$, and $90^\circ$. The $60^\circ$ case is in particular interesting, as it simulates the observational capabilities for L$_4$/L$_5$ missions, which are currently under discussion to optimize space weather predictions \citep{gop11}.

\section{MHD Simulations}
\label{s:models}

Simulations give us the opportunity to study the kinematical behavior of ICMEs from an ideal location as observer, namely top view along the Sun-Earth line, which enables us to directly measure the time-distance profile of the ICME evolution. This is in contrast to how we actually observe ICMEs, \textit{i.e.}\ from a side view involving LoS and scattering effects, from which we measure an elongation angle that needs to be converted into radial distance by using different conversion methods. From the model we obtain both views, and use these two capabilities to test the efficiency of the CHM method to derive the kinematics of an ICME measured from synthetic Jmaps as observed by STEREO-A and to calculate ICME propagation directions. Synthetic Jmaps are produced from three different observing positions, \textit{i.e.}\ $30^\circ$, $60^\circ$, and $90^\circ$ between ICME apex and remote spacecraft in order to test the influence of the observing angle to the resulting directions and kinematics.

The numerical modeling of the simulated ICME is done by using the Space Weather Modeling Framework \citep[SWMF:][]{tot05}.
Along the Sun-Earth line the grids of the model vary in spatial resolution from $0.5$ R$_\odot$ close to the Sun to $0.25$ R$_\odot$ in the middle and $0.125$ R$_\odot$ close to Earth. Synthetic \textit{in-situ} data, \textit{i.e.}\ magnetic field, plasma density and plasma bulk velocity, are produced for the location of Earth. This is done using the semiempirical MHD model developed by \citet{coh07}, which applies the empirical Wang--Sheeley--Arge model \citep[WSA:][]{was90,arp00} and a flux rope CME initiated at the solar surface \citep[similar setup as in][]{lug09}. During the last 16 hours before the shock hits Earth the temporal resolution is about 2\,--\,3 minutes in order to capture the details of the transition of the shock.

The modeled fictitious ICME is launched on 15 February 2011 01:00:00 into a steady state background solar wind ($V_{\rm sw}\approx 430$ km s$^{-1}$). It has a very wide angular extent within the ecliptic and is expanding to all sides. It is directed toward the virtual \textit{in-situ} spacecraft at the position of Earth and the ICME apex arrives 48 hours later on 17 February 2011 01:00:00. The ICME is faster than the ambient solar wind. It decelerates from 1100 km s$^{-1}$ at 60 R$_\odot$ to 800 km s$^{-1}$ at 150 R$_\odot$, and the shock finally arrives with $\approx 650$ km s$^{-1}$ at the \textit{in-situ} spacecraft.

\begin{figure} 
 \centerline{\includegraphics[width=\textwidth,height=\textheight,keepaspectratio]{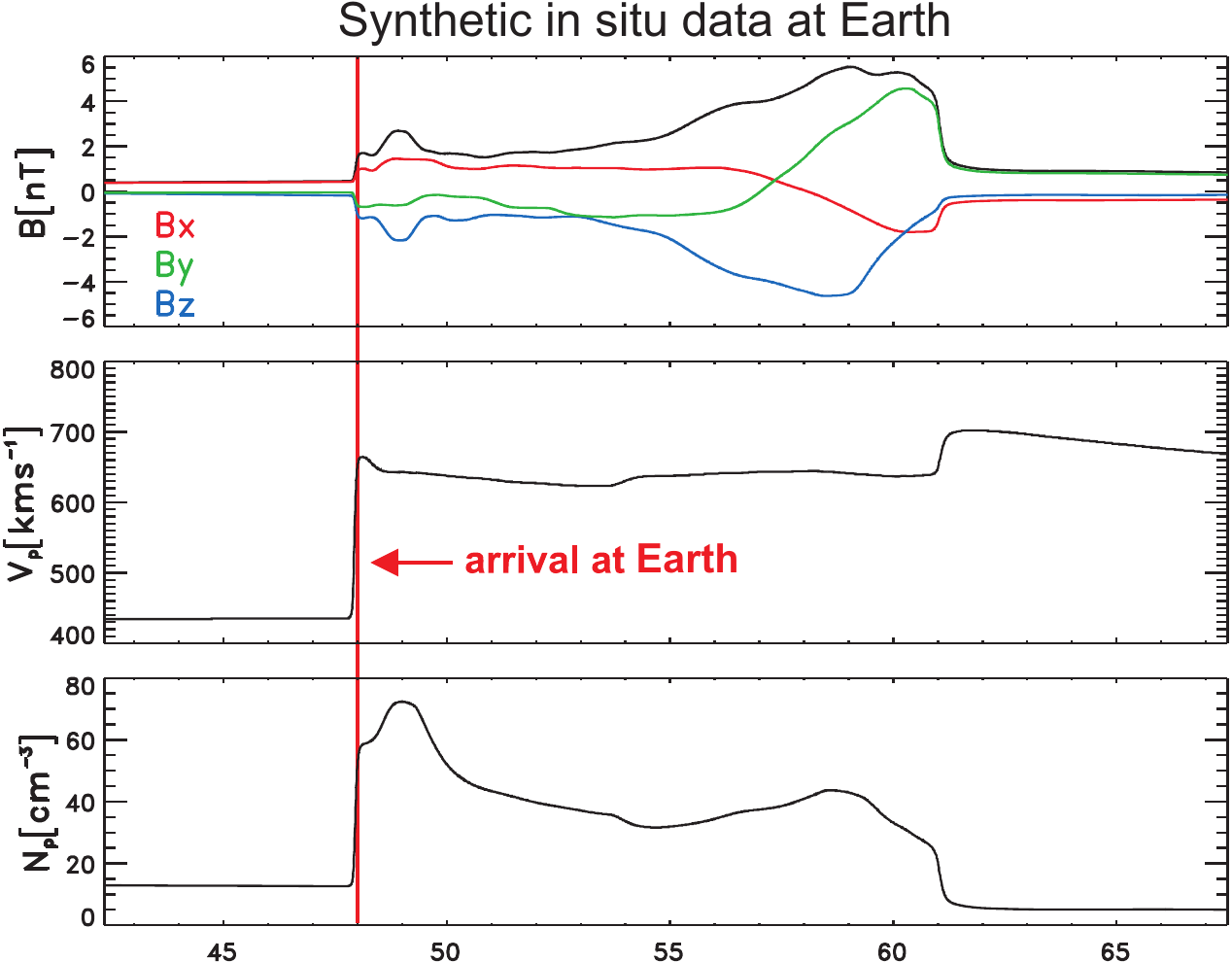}}
 \caption{Synthetic \textit{in-situ} data at the position of Earth. The upper panel shows the magnetic field with its three components (cartesian coordinate system) and the total magnetic field strength, the middle panel shows the speed and the bottom panel the proton number density. The vertical red line marks the shock arrival time at Earth that is used for further calculations. The \textit{x}-axis gives the time in hours from launch of the CME at the Sun.}
 \label{fig:insitu}
\end{figure}

Figure \ref{fig:insitu} shows the synthetic \textit{in-situ} data extracted at the position of Earth. In the upper panel the three components of the magnetic field and the total magnetic field strength (black line) are plotted in cartesian coordinates with origin in Sun-center, \textit{x}- and \textit{y}-axes lie in the ecliptic plane and \textit{z}-axis points to the ecliptic north pole to complete the right hand triad. The middle and the bottom panel show the speed in km~s$^{-1}$ and the proton number density in cm$^{-3}$, respectively. The red vertical line indicates the arrival of the shock preceding the ICME. If the shock is absent, the arrival time cannot be defined unambiguously. This could yield an additional uncertainty.

\begin{figure} 
 \centerline{\includegraphics[width=\textwidth,keepaspectratio]{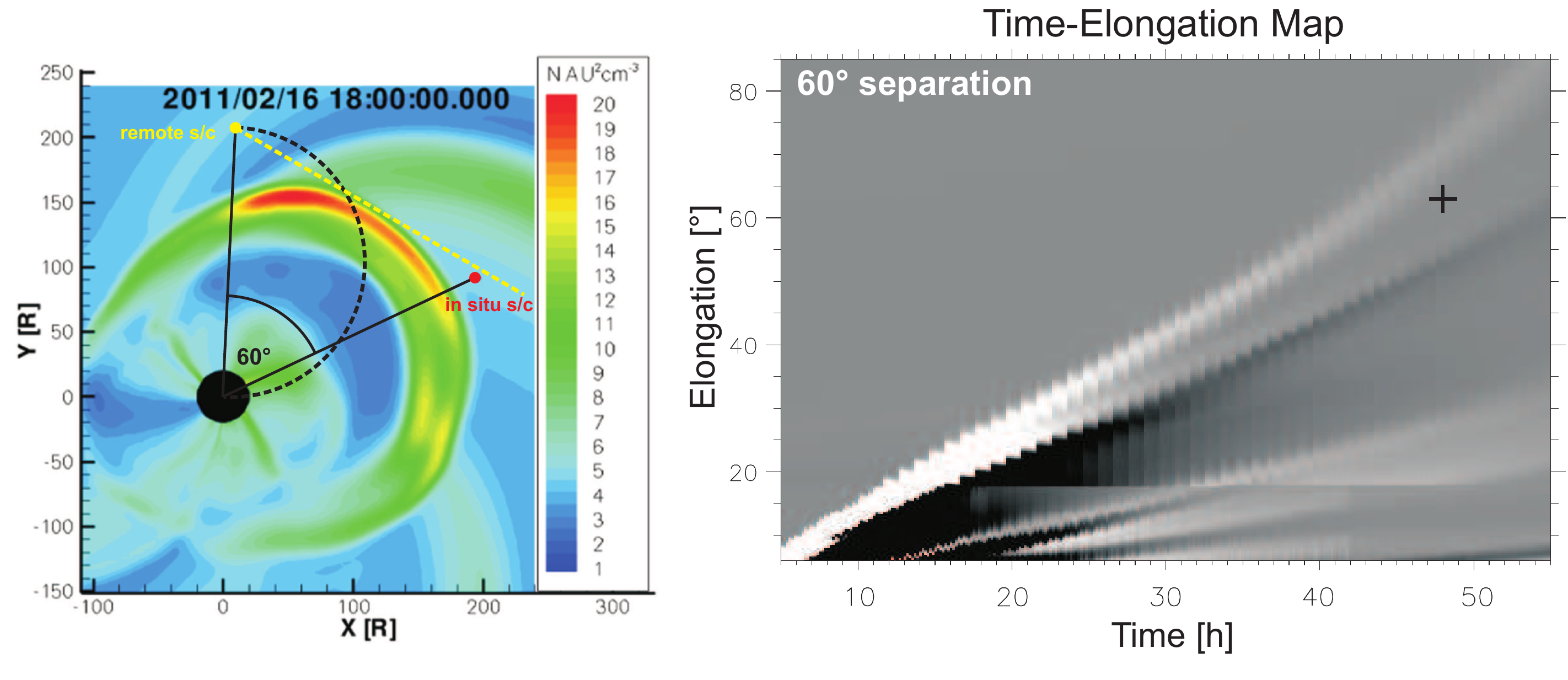}}
 \caption{Left: top view of the modeled ICME for $60^\circ$ separation between remote (yellow dot) and \textit{in-situ} spacecraft (red dot). The black semicircle is the Thomson surface and the dashed yellow line is the LoS along the ICME front. Right: side view of the propagating ICME simulated as observed by STEREO-A/HI. The bright track shows the evolution in time of the apparent ICME front. The black cross marks the elongation of the \textit{in-situ} spacecraft at shock arrival time.}
 \label{fig:topjmap}
\end{figure}

The left part of Figure \ref{fig:topjmap} shows the location of the observing remote spacecraft (yellow dot) and \textit{in-situ} spacecraft (red dot) as viewed from top at one time step of the ICME evolution. The black semicircle is the Thomson surface \citep[TS:][]{vou06,how12} and the dashed yellow line is the LoS along the ICME front, \textit{i.e.}\ the projection of the TS onto the ecliptic plane, from the view of the remote spacecraft for which the Jmap is produced. The corresponding Jmap is shown on the right side of Figure \ref{fig:topjmap}. The bright features are the integrated signal along the LoS with its intensity maximum at the TS. The method to produce synthetic Jmaps is described in \citet{lug09}.

\section{Analysis Methods}
\label{s:methods}
To determine the kinematical properties of an ICME, we usually measure the apparent front, either in the direct white-light images or within Jmaps. Tracking the bright features in a Jmap has two advantages. 
First, the contrast between dense and less dense areas can easily be seen, which enables us to track ICMEs over larger elongations, hence, radial distances. Especially in HI2 images the brightness rapidly decreases and it is hardly feasible to identify the front at larger elongations. Second, due to the shape of the track that is directly visible in the Jmap, \textit{i.e.}\ the evolution of elongation with time, it is possible to get a first estimate of the direction of the ICME \citep{she08}. This is used in fitting methods as the Fixed-$\phi$ fitting \citep{kah07,rou08} or the Harmonic Mean fitting methods \citep{lug10,moe11} that can also be used to predict ICME arrival times. However, these fitting methods use the assumption of constant speed, which is very useful to get a rough estimate for the direction and the arrival time of the ICME. But to analyze variations in the kinematical behavior due to interactions with \textit{e.g.}\ high speed streams, corotating interaction regions or preceding ICMEs \citep[\textit{e.g.}][]{gop01a,bur02,lug09b,tem11,tem12} it is required to apply a method, which allows the ICME speed to vary during propagation.

Such an approach, which uses additional \textit{in-situ} data to constrain measurements from white-light data, is described in \cite{rol12}. In this study we validate the CHM method (using the constraint of arrival time) and derive the effective error in the ICME distance--time and speed--distance profiles.

\subsection{Constrained Harmonic Mean Method}

In the following we briefly outline the CHM method and its main features, for more details we refer the interested reader to \citet{rol12}. At this point we stress that the CHM method is not a forecasting method but an approach to analyze ICMEs after they are remotely observed and measured \textit{in-situ}, respectively. In this way we are able to derive a kinematical profile for the purpose of analyzing variations in speed. In order to convert the measured elongation angle [$\epsilon(t)$] into a unit of distance we use an approach assuming a circular ICME front, the so-called Harmonic Mean method \citep[HM:][]{howtap09,lug09b}. HM was shown to be a quite good approximation for very wide events. This method assumes the observer to look along the tangent to the spherical ICME front and more or less disregards the concept of the intensity maximum on the TS. Equation (\ref{eq:hmean1}) shows the conversion of $\epsilon(t)$ into the radial distance of the ICME apex [$R_{\rm HM}(t)$] for a given propagation direction [$\phi$]:

\begin{equation}
R_{\rm HM}(t) = \frac{2 d_{\rm o} \sin \epsilon(t)}{1 + \sin (\epsilon(t) + \phi)},
\label{eq:hmean1}
\end{equation}
where $d_{\rm o}$ is the Sun-observer distance. For the constraint with the \textit{in-situ} arrival time we need the distance of the ICME front in the direction of the \textit{in-situ} spacecraft. This is given by

\begin{equation}
R_{\rm HMi,t_a}(\phi) = R_{\rm HM,t_a}(\phi)\cos{\delta},
\label{eq:cosinus}
\end{equation}
where $R_{\rm HMi,t_{a}}(\phi)$ is the distance of the ICME front in the direction of the \textit{in-situ} spacecraft at arrival time [$t_{\rm a}$], $R_{\rm HM,t_{a}}$ is the distance of the ICME apex at arrival time and $\delta$ is the angle between the line of the \textit{in-situ} spacecraft to the Sun [$d_{\rm i}$], and $R_{\rm HM}(t)$. To determine the propagation direction we now calculate 

\begin{equation}
\Delta d_{\rm HMi}(\phi) = R_{\rm HMi,t_{a}}(\phi) - d_{\rm i},
\label{eq:rinsitu}
\end{equation}
for $\phi \in [0^{\circ},180^{\circ}]$. The direction providing the minimum value for $\Delta d_{\rm HMi}(\phi)$ is the resulting propagation direction $\phi$ that is now used to convert $\epsilon(t)$ into distance by using Equation (\ref{eq:hmean1}).

In other words: the input parameter $\phi$, and therefore also the shape of the distance--time profile, is changed until the best accordance with the \textit{in-situ} arrival time is found. This ensures that the HI observations are consistent with the \textit{in-situ} data, which serve as a boundary condition for the kinematical profiles. In addition, the input paramter $\phi$, used for the best agreement, is the calculated propagation direction of the ICME apex \citep{moe09jgr,moe10,rol12}.

\subsection{Speed Calculation}

\begin{figure} 
 \centerline{\includegraphics[width=\textwidth,height=\textheight,keepaspectratio]{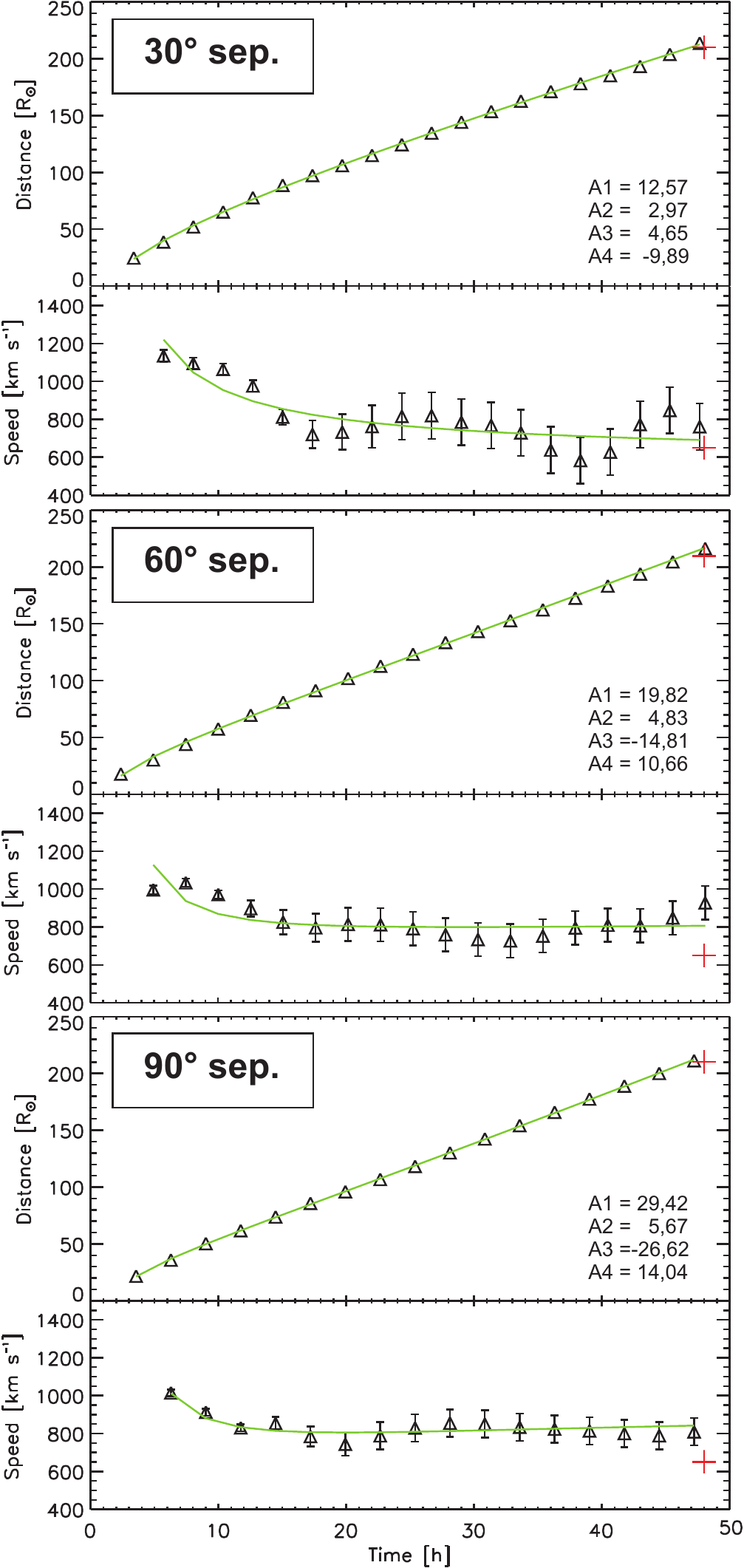}}
 \caption{Distance--time profiles with over plotted fits for $30^\circ$, $60^\circ$, and $90^\circ$ separation of remote and \textit{in-situ} spacecraft. In each case the upper panel shows the distance measurements converted from the elongation--time profiles (triangles), the over plotted fits and the fitting parameters in R$_\odot$. The lower panels show the time derivatives from the distance--time profiles and the over plotted time derivatives of the distance--time fits. The red crosses mark the \textit{in-situ} measured arrival time and speed, respectively.}
 \label{fig:fits}
\end{figure}

When the ICME speed profile is derived from the calculated distance its shape strongly depends on the uncertainties in the distance profile. Therefore it is necessary to fit the distance profile and use this fit for further calculations. In \citet{gal03} and \citet{bei11} fitting functions and methods are described for coronagraphic FoVs close to the Sun. These functions take into account the strong acceleration and deceleration at the beginning of the CME evolution \citep{zha01}. Since we analyze data from HI we are mainly interested in the propagation phase, during which the ICME adjusts to the solar wind speed, and developed a function, which takes into account the different parts of the distance--profile for the HI FoV:

\begin{equation}
F(x)=A_{1}\ln(x)+A_{2}x+A_{3}\sqrt{x}+A_{4},
\label{eq:fit}
\end{equation}
where the first term reproduces the strong deceleration at the beginning of the HI1 FoV, the second term describes the roughly constant speed during the propagation phase. The third term takes into account the slowly decreasing/increasing speed and is needed to avoid an overestimation of the arrival speed at 1~AU, and the constant is the \textit{y}-intercept. Figure \ref{fig:fits} shows the distance--time fits and the fitting parameters for all three separations. The time derivatives of the distance--time fits are over plotted on the speed--time profiles, derived from the distance measurements converted from the elongation--time profiles. The error bars show the standard deviations calculated from measuring each track five times.

\subsection{Error Analysis}

Uncertainties in the analysis of ICME kinematics and directions are due to various reasons. By using a modeled ICME it is possible to provide an error estimation for the most significant contributions.

If the apparent front of an ICME is measured from white-light images the intensity range is varying, \textit{i.e.}\ if a track is measured by more than one person there will be a certain spread of measurements. To quantify this error we determine an error range for the measurements by performing two surveys---one very high (outermost) and one very low (innermost) track (see Section \ref{s:kin}). The mean differences between these two tracks lie between $\pm 0.3^\circ$ (HI1) and $\pm 1.35^\circ$ elongation (HI2) for all three separation cases.

To estimate the error due to visually tracking of the elongation--time profile, the Jmap track is measured five times. Out of this sample we calculate the mean values and standard deviations for every Jmap track. The elongation errors for all three separation cases lie in a range of about $\pm 0.1^\circ$ (HI1) and $\pm 0.4^\circ$ (HI2). 

Table \ref{tab:error} lists the mean values for the three cases of different separation angles between \textit{in-situ} and remote spacecraft and both HI instruments, for the two derived errors in elongation, \textit{i.e.}\ the intensity range, where measurements could be taken within the Jmap and the measurement error itself. For further calculations only the measurement error is taken into account. Table \ref{tab:error2} gives the errors for the resulting distance in R$_{\odot}$ and the speed in km~s$^{-1}$.

We only consider errors due to the measurements itself. Uncertainties beacause of the assumed circular shape or the angular width of the ICME front are not taken into account. Since HM assumes a width of $180^\circ$ it may not be the right approach for narrow events. In this case the F$\phi$ method should provide more reliable results.

\begin{table}[!htbp]
\caption{Summary of calculated errors denoted in elongation for each separation case and each heliospheric instrument: HI1 and HI2. The first row lists the standard deviation of the innermost and the outermost Jmap track, \textit{i.e.}\ the possible Jmap tracking area. The second row shows the measurement error derived out of the five tracks in elongation.}
\label{tab:error}
\begin{tabular}{lcccccc}
\cmidrule{2-7}
 & \multicolumn{2}{c}{$30^\circ$ separation}& \multicolumn{2}{c}{$60^\circ$ separation} & \multicolumn{2}{c}{$90^\circ$ separation}\\
\cmidrule{2-7}
  & HI1 & HI2 & HI1 & HI2 & HI1 & HI2\\
\hline
 Jmap tracking area & $\pm 0.25^\circ$ & $\pm 1.35^\circ$ & $\pm 0.3^\circ$ & $\pm 1^\circ$ & $\pm 0.3^\circ$ & $\pm 0.7^\circ$\\
 mean standard deviation & $\pm 0.1^\circ$ & $\pm 0.4^\circ$ & $\pm 0.5^\circ$ & $\pm 1.9^\circ$ & $\pm 0.5^\circ$ & $\pm 1.4^\circ$\\
\hline\\
\end{tabular}
\end{table}

\begin{table}[!htbp]
\caption{Resulting errors for the ICME kinematics. The first row lists the distance error in R$_{\odot}$ based on the mean standard deviations for both instruments. The second row shows the speed error derived from the standard deviations.}
\label{tab:error2}
\begin{tabular}{lcccccc}
\cmidrule{2-7}
 & \multicolumn{2}{c}{$30^\circ$ separation}& \multicolumn{2}{c}{$60^\circ$ separation} & \multicolumn{2}{c}{$90^\circ$ separation}\\
\cmidrule{2-7}
  & HI1 & HI2 & HI1 & HI2 & HI1 & HI2\\
\hline
 distance error {[}R$_{\odot}${]} & $0.4$ & $1.7$ & $0.4$ & $1.6$ & $0.4$ & $1.3$\\
 \hline
 speed error {[}km~s$^{-1}${]} & $31$ & $128$ & $34$ & $102$ & $27$ & $81$\\
\hline\\
\end{tabular}
\end{table}

\section{Results}

In this section we provide the calculated directions and the kinematical profiles derived from the measurements within the synthetic Jmaps by using the CHM method. For validation, these results are compared to the ``true'' apex kinematics, measured from the top view of the modeled ICME.

\subsection{Kinematics and Direction}
\label{s:kin}

\begin{figure} 
 \centerline{\includegraphics[width=\textwidth,height=\textheight,keepaspectratio]{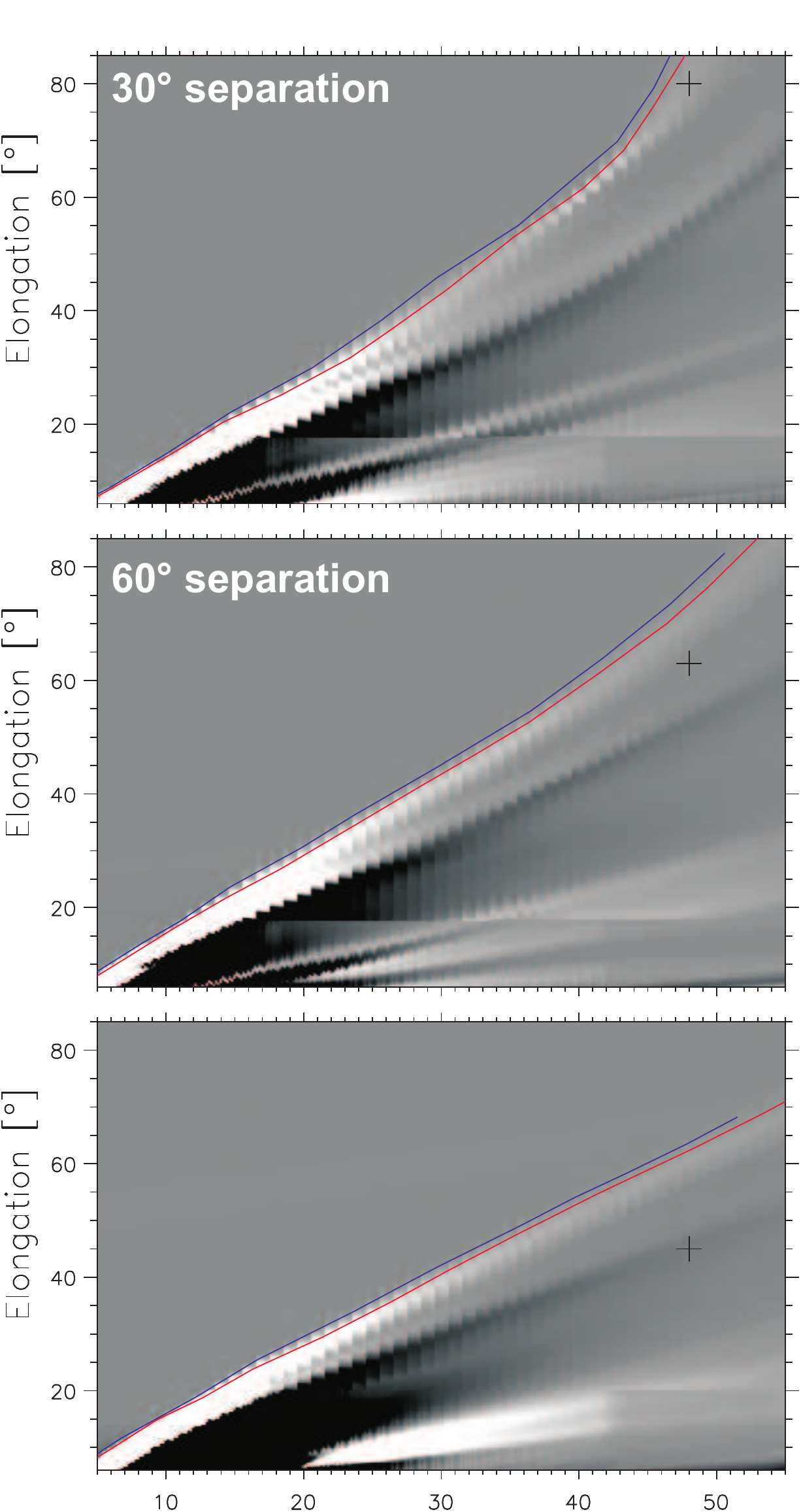}}
 \caption{Jmaps for the three separations with the measured tracks of the ICME front over plotted. The red line is the innermost track and the blue line is the outermost track. The black crosses mark the elongation of the \textit{in-situ} spacecraft at shock arrival time.}
 \label{fig:jmaptrack}
\end{figure}

Due to the fact that the ICME is directed toward the virtual \textit{in-situ} spacecraft, that is located at the position of Earth, the real ICME direction is E0/W0. For all three separation cases the Jmaps with the outermost and innermost tracks are shown in Figure \ref{fig:jmaptrack}. The red line is the innermost and the blue line is the outermost edge of the ICME front. The black cross marks the elongation of the \textit{in-situ} spacecraft at shock arrival time. Because of the large angular width of the ICME, LoS effects make the ICME to appear earlier at 1~AU than actually measured \textit{in-situ}.

\begin{figure} 
 \centerline{\includegraphics[width=\textwidth,height=\textheight,keepaspectratio]{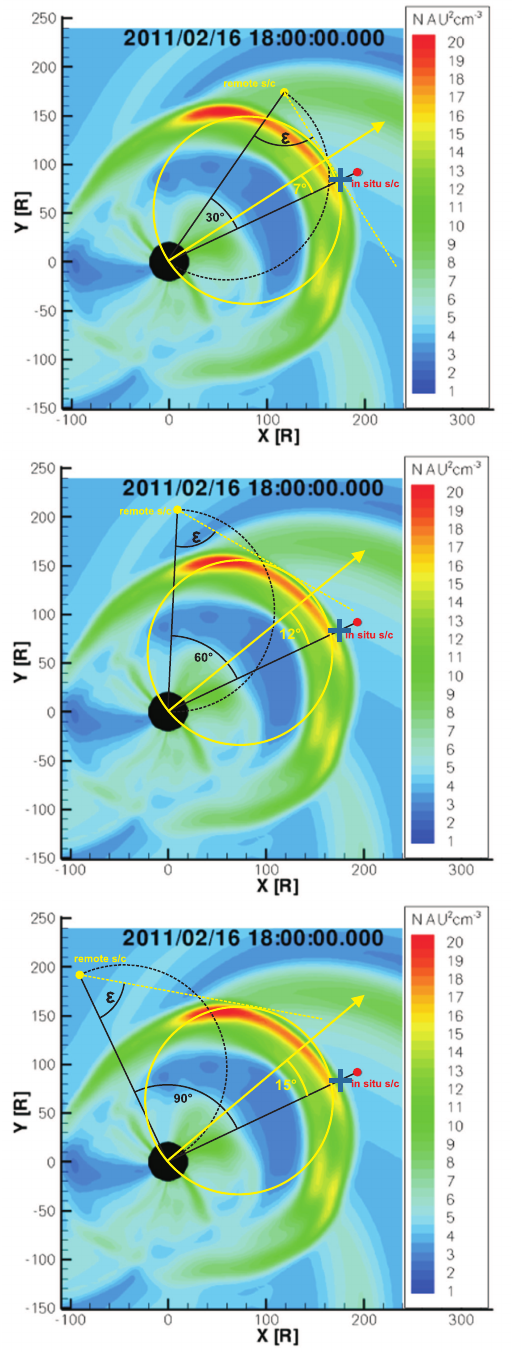}}
 \caption{Top view of the modeled ICME. The denser region (red part) is located at the western flank of the ICME. The spacecraft positions (yellow and red dots) and resulting directions from the CHM method (yellow arrow) for all three separations (top: $30^\circ$, middle: $60^\circ$, bottom: $90^\circ$) are over plotted. The black semicircle is the TS and the yellow dashed line is the tangent to the circle assumed by the Harmonic Mean approximation. The blue cross marks the ICME apex that was measured from top view.}
 \label{fig:topview}
\end{figure}
The resulting directions obtained for the mean of the innermost and the outermost Jmap tracks are shown in Figure \ref{fig:topview}. The yellow circle is the assumed circular shape of the HM approximation and the yellow arrow marks the direction of the ICME apex. The dashed semicircle is the TS. For all three separation cases the resulting directions are consistent within a few degrees ($30^\circ$ sep: $\phi=$ W7; $60^\circ$ sep: $\phi=$ W12; $90^\circ$ sep: $\phi=$ W15; true dir.: E0/W0). Because of the corotating interaction region (CIR), heading in front of the ICME, the western flank is denser, which may be the reason for a result toward the western direction. The resulting propagation angles for the innermost tracks are 2\,--\,$4^\circ$ larger than the directions calculated from the outermost tracks (see Table \ref{tab:direc}). To demonstrate how well the CHM method works, we calculate the directions also for the constrained F$\phi$ method using the constraint of the \textit{in-situ} data point of the arrival time of the ICME \citep[see][]{rol12}. F$\phi$ assumes a single point moving radially away from the Sun and is thus more dependent on the separation of both spacecraft and/or the remote spacecraft to the ICME apex than HM.

\begin{figure} 
 \centerline{\includegraphics[width=\textwidth,clip=]{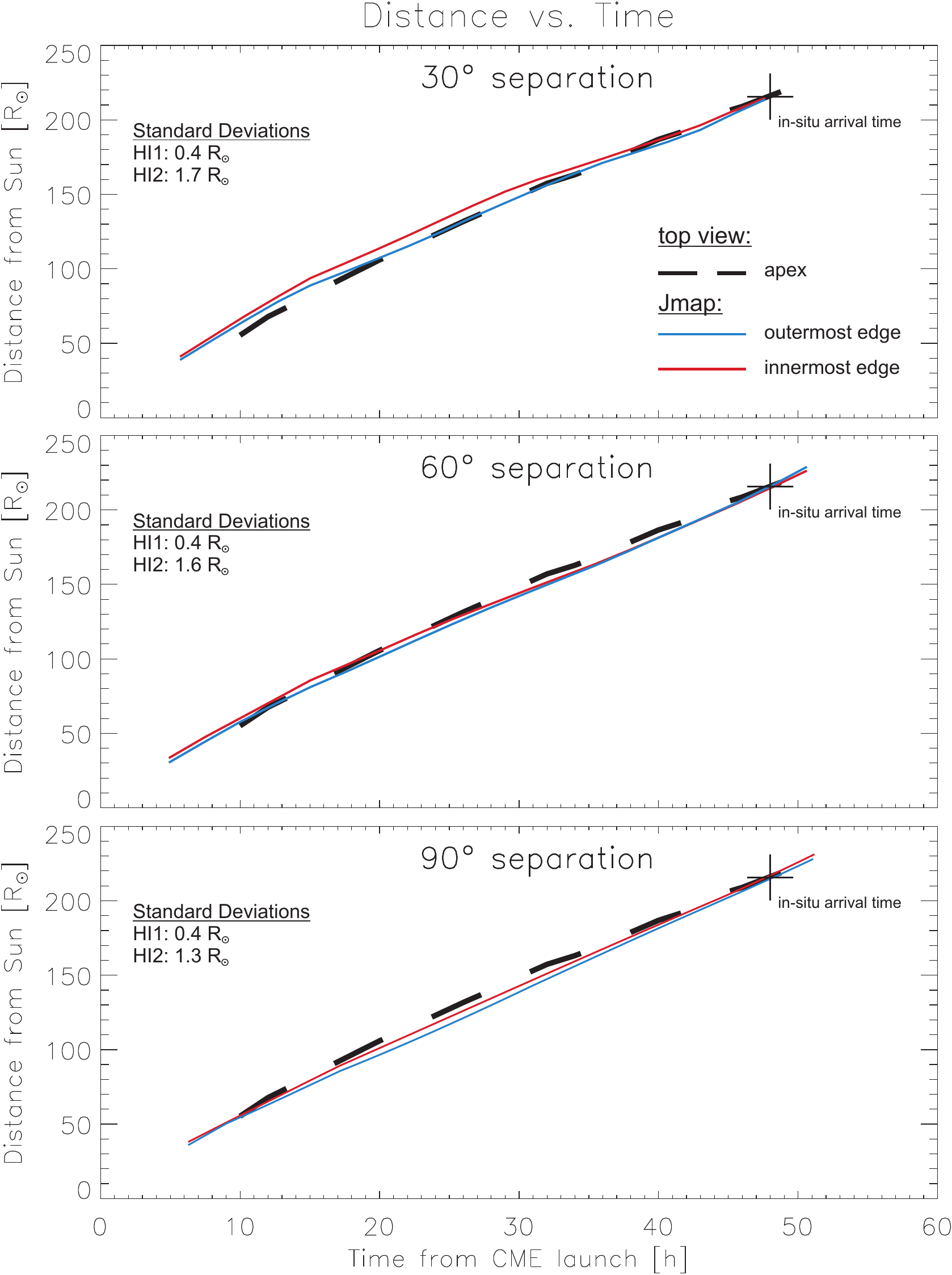}}
 \caption{Distance--time profiles for $30^\circ, 60^\circ$, and $90^\circ$ separation from the \textit{in-situ} spacecraft. The dashed line shows the top view measurement of the apex, the solid lines are the conversions from the innermost (red) and the outermost (blue) Jmap measurements., the shaded areas are the error ranges. The crosses indicate the arrival time at the \textit{in-situ} spacecraft.}
 \label{fig:dist}
\end{figure}

\begin{table}[!htbp]
\caption{Resulting ICME propagation directions for the three separation cases. Constrained Fixed-$\phi$ (CF$\phi$) and constrained Harmonic Mean (CHM) methods are used to calculate the directions for the innermost (inn.) as well as for the outermost (out.) track in the Jmap.}
\label{tab:direc}
\begin{tabular}{lcccccc}
\cmidrule{2-7}
 & \multicolumn{2}{c}{$30^\circ$ separation}& \multicolumn{2}{c}{$60^\circ$ separation} & \multicolumn{2}{c}{$90^\circ$ separation}\\
\cmidrule{2-7}
  & inn. & out. & inn. & out. & inn. & out.\\
\hline
 CHM & W6 & W8 & W10 & W14 & W15 & W16\\
 CF$\phi$ & W13 & W17 & W22 & W28 & W32 & W25\\
\hline\\
\end{tabular}
\end{table}
In Figure \ref{fig:dist} the converted Jmap measurements (solid lines) are compared to the real ICME apex distance--time profile (dashed line). The top panel shows the $30^\circ$ separation case which overestimates the distance especially for the HI1 FoV. The best match of measurements and ICME apex is given for the innermost edge (red line) in the Jmap. The middle panel shows the distance--time profile for $60^\circ$ separation revealing that both tracks are in a relatively good agreement with the ICME apex compared to the other separation cases. The bottom panel shows the $90^\circ$ separation case. In contrast to the $30^\circ$ separation case, the distance is underestimated most of the time. For $60^\circ$ and $90^\circ$ separation the measurements taken at the outermost edge of the Jmap track reveal the best agreement to the apex measurements.

\begin{figure} 
 \centerline{\includegraphics[width=\textwidth,height=\textheight,keepaspectratio]{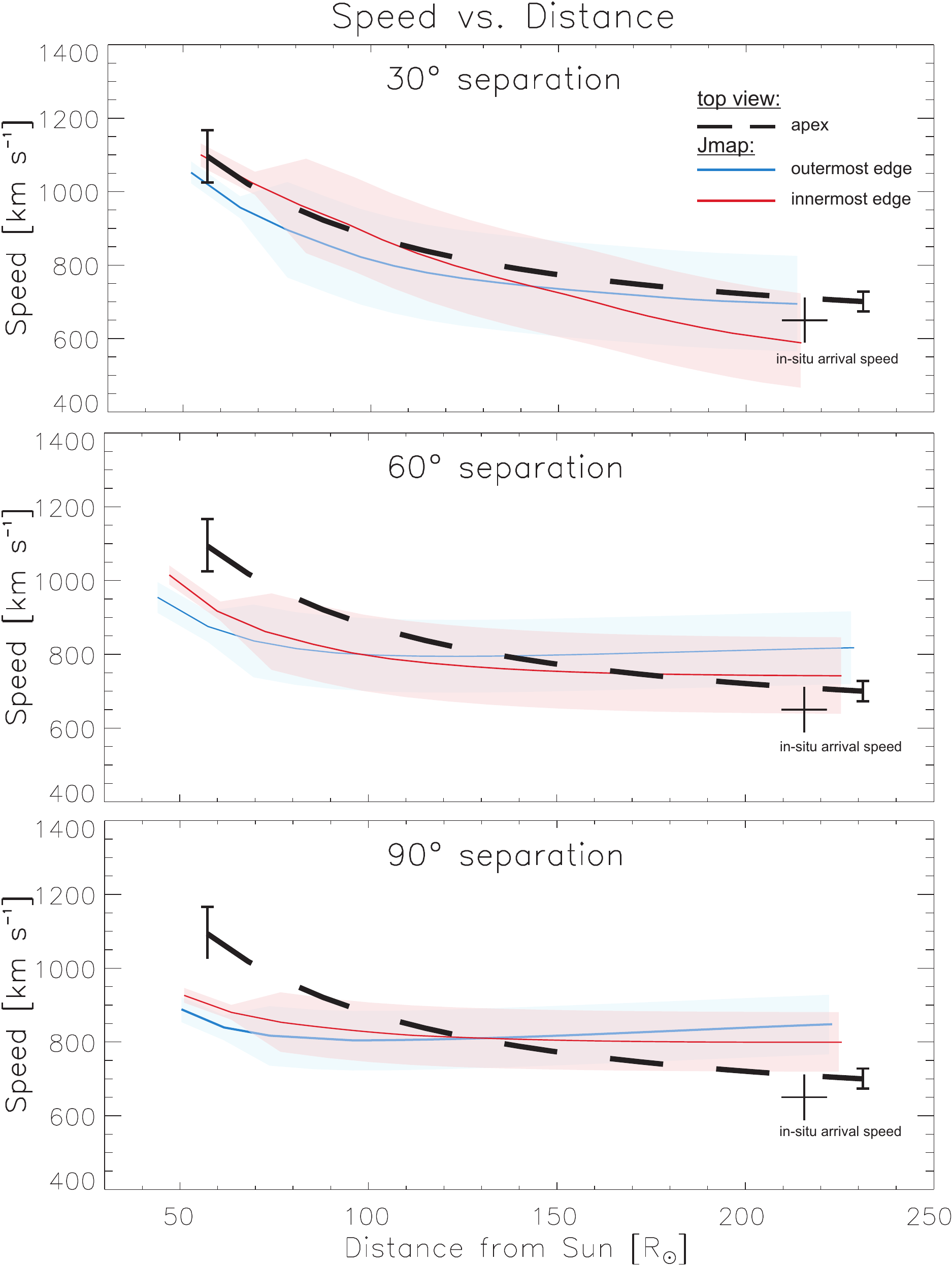}}
 \caption{Speed--distance profiles for $30^\circ, 60^\circ$, and $90^\circ$ separation from the \textit{in-situ} spacecraft. The dashed line shows the time derivatives from the top view distance measurement of the apex, the solid lines are the time derivatives from the conversions from the innermost (red) and the outermost (blue) Jmap measurements. The crosses indicate the arrival speed at the \textit{in-situ} spacecraft.}
 \label{fig:speed}
\end{figure}

Figure \ref{fig:speed} displays the same approach as Figure \ref{fig:dist} but for the speed--distance profile. The ICME apex enters the HI1 FoV with a speed of $\approx 1100$ km~s$^{-1}$. It slowly decreases to a mean speed of $\approx 750$ km~s$^{-1}$. The ``true'' speed profile of the ICME apex is derived from the top view images. With growing distance from the Sun its standard deviation decreases from 148 km~s$^{-1}$ to about 62 km~s$^{-1}$ because of the increasing resolution of the simulation. The speeds derived from the innermost Jmap tracks for all three separation cases show a stronger deceleration than the speed profiles from the outermost edges. The $30^\circ$ separation case reveals the best agreement, while the $60^\circ$ as well as the $90^\circ$ observing angle in the early propagation phase under- and in the late propagation phase overestimate the ICME speed. With growing separation between remote spacecraft and ICME apex the overestimation of the arrival speed increases. While in the $30^\circ$ separation case the arrival speed of the outermost edge is quite close to the apex speed, the innermost track underestimates the arrival speed by about 100 km s$^{-1}$. From both other viewing angles the arrival speed is overestimated on average by about 75 km s$^{-1}$ ($60^\circ$ separation) and 125 km s$^{-1}$ ($90^\circ$ separation), respectively. As well as for the distance--profile, measuring the Jmap track at the innermost edge in the $30^\circ$ and $60^\circ$ separation cases gives more consistent results compared to the apex.

\section{Discussion and Conclusions}

One aim of this study was to validate the constrained Harmonic Mean method (CHM) introduced in \citet{rol12}. This method assumes a spherical ICME front always attached to the Sun and the observer to look along the tangent of the front. The CHM method uses single-spacecraft heliospheric imager (HI) observations and \textit{in-situ} data at 1~AU. By constraining the kinematics with the \textit{in-situ} arrival time it is possible to derive the propagation direction for the ICME apex. In addition we can provide distance--time and speed--distance profiles for the entire propagation phase. The second purpose was to provide a reliable error estimate for the kinematics of the ICME when applying the CHM method to Jmap measurements.

In order to test the CHM method we used a numerical simulation of an ICME and analyzed synthetic Jmaps from three different vantage points ($30^\circ$, $60^\circ$, and $90^\circ$ separation of remote spacecraft and \textit{in-situ} spacecraft). The resulting directions and kinematics for each of the three separation angles were compared to the ``true'' kinematics of the ICME apex, \textit{i.e.}\ free from effects due to Thomson scattering or conversion methods, derived from the top view of the modeled ICME.

The directions obtained seem to be not very sensitive to the separation angle between remote and \textit{in-situ} spacecraft ($30^\circ$ sep: $\phi=$ W7; $60^\circ$ sep: $\phi=$ W12; $90^\circ$ sep: $\phi=$ W15; true dir.: E0/W0). The density of the ICME is higher in its western flank as a result of a CIR ahead of the ICME. This fact could have influenced our results for the propagation direction and shifted them to a western direction. Another possible reason for that could be that the CHM method assumes a smaller curvature radius of the ICME front than shown by the simulated ICME. It still has to be tested, how the results of the CHM method depend on the ecliptic extend, \textit{i.e.}\ the width of the ICME.

Future studies should prove the behavior of the method when ICMEs with different widths are investigated. In \citet{lug11} the results of the F$\phi$ and HM fitting methods, which assume constant speed, were tested by analyzing a simulated ICME from different viewing angles. Consistent with our study, a stronger dependance of the observing angle for F$\phi$ than for HM was obtained when calculating the propagation direction.

The ICME kinematics were compared to the actual distance--time and speed--distance profiles. The best agreement for the distance is obtained for the $60^\circ$ separation case. The $30^\circ$ separation case overestimates the distance most of the time while in the $90^\circ$ separation case the distance is too small compared to the ICME apex derived from the top view. For the speed--distance profiles the best accordance is revealed by the $30^\circ$ separation case. Both other settings first under- and then overestimate the speed. The differences between the true and the calculated arrival speed increases with growing separation ($30^\circ$ sep: $\Delta V_{\rm arr}\approx-50$ km s$^{-1}$, $60^\circ$ sep: $\Delta V_{\rm arr}\approx+75$ km s$^{-1}$, $90^\circ$ sep: $\Delta V_{\rm arr}\approx+125$ km s$^{-1}$). In general, conversion and fitting methods making special geometrical assumptions on the shape of the ICME front, \textit{e.g.}\ F$\phi$, HM or SSE methods, overestimate the arrival speed with increasing observation angle.

The error estimation was done by taking into account two different error sources. The intensity area in the Jmap, where the apparent ICME front could be measured, varies between $\pm 0.3^\circ$ (HI1) and $\pm 1.35^\circ$ elongation (HI2). This should reflect the range of tracks if more than one person measures the front. This is in agreement with the results found by \citet{wil09} who quantified the elongation range of chosen tracks measured within a Jmap with less than $\pm1^\circ$ for HI1 and $\pm2^\circ$ for HI2. We find that in general, measuring the Jmap track at the innermost edge gives more consistent results compared to the apex kinematics. The second error is due to the visual measuring of the front and is obtained by calculating the standard deviations out of five measurements taken that lie in a range of $\pm 0.1^\circ$ (HI1) and $\pm 0.4^\circ$ (HI2). The errors within the speed-profiles lie in a range of 27--34 km s$^{-1}$ (HI) and 81--128 km s$^{-1}$ (HI2).

The two STEREO spacecraft already are on the backside of the Sun, and thus a forecast for Earth directed ICMEs using the heliospheric imagers is no longer possible. To improve existing space weather forecast an L$_4$/L$_5$ mission carrying heliospheric imagers is a valuable option. The side view of the entire distance from the Sun up to 1~AU is a big advantage upon coronagraphic observations from Earth view.

%


%
 \begin{acks}
This work has received funding from the European Commission FP7 Project n$^\circ$ 263252 [COMESEP]. M.~T.\ acknowledges the Austrian Science Fund (FWF): V195-N16. This research was supported by a Marie Curie International Outgoing Fellowship within the 7th European Community Framework Programme. N.~L.\ was supported by NSF AGS1239704 and NASA NNX12AB28G. Simulation results were obtained using the Space Weather Modeling Framework, developed by the Center for Space Environment Modeling, at the University of Michigan with funding support from NASA ESS, NASA ESTO-CT, NSF KDI, and DoD MURI. We thank the STEREO SECCHI/IMPACT/PLASTIC teams for their open data policy.
 \end{acks}


%
%
 \bibliographystyle{spr-mp-sola}
%

\end{article}
\end{document}